\documentclass{article}

\usepackage{booktabs}
\usepackage{graphics}
\usepackage{graphicx}
\usepackage{listings}
\usepackage{nameref}
\usepackage{algorithm}
\usepackage{algpseudocode}
\algrenewcommand\algorithmicrequire{\textbf{Input:}}
\algrenewcommand\algorithmicensure{\textbf{Output:}}
\usepackage{bm}
\usepackage{amsmath}
\usepackage{amssymb}
\usepackage{caption} 
\usepackage{multirow}
\usepackage{arxiv}

\usepackage[utf8]{inputenc} 
\usepackage[T1]{fontenc}    
\usepackage{hyperref}       
\usepackage{url}            
\usepackage{booktabs}       
\usepackage{amsfonts}       
\usepackage{nicefrac}       
\usepackage{microtype}      
\usepackage{lipsum}
\usepackage{graphicx}
\graphicspath{ {./images/} }

\title{Bacteriophage classification for assembled contigs using Graph Convolutional Network}

\author{
 Jiayu Shang \\
  Dept. of Electrical Engineering\\
  City University of Hong Kong\\
  Kowloon, Hong Kong SAR, China\\
  \texttt{jyshang2-c@my.cityu.edu.hk}  \\
   \And
 Jingzhe Jiang \\
   Ministry of Agriculture\\
   South China Sea Fisheries Research Institute\\
   Chinese Academy of Fishery Sciences\\
   Guangzhou, Guangdong Province, China\\
  \texttt{jingzhejiang@scsfri.ac.cn} \\
  \And
 Yanni Sun \\
  Dept. of Electrical Engineering\\
  City University of Hong Kong\\
  Kowloon, Hong Kong SAR, China\\
  \texttt{yannisun@cityu.edu.hk} \\
}

\begin{document}
\maketitle
\begin{abstract}
\textbf{Motivation:} Bacteriophages (aka phages), which mainly infect bacteria, play key roles in the biology of microbes. As the most abundant biological entities on the planet, the number of discovered phages is only the tip of the iceberg. Recently, many new phages have been revealed using high throughput sequencing, particularly metagenomic sequencing. Compared to the fast accumulation of phage-like sequences, there is a serious lag in taxonomic classification of phages. High diversity, abundance, and limited known phages pose great challenges for taxonomic analysis. In particular, alignment-based tools have difficulty in classifying fast accumulating contigs assembled from metagenomic data. \\
\textbf{Results:} In this work, we present a novel semi-supervised learning model, named PhaGCN, to conduct taxonomic classification for phage contigs. In this learning model, we construct a knowledge graph by combining the DNA sequence features learned by convolutional neural network (CNN) and protein sequence similarity gained from gene-sharing network. Then we apply graph convolutional network (GCN) to utilize both the labeled and unlabeled samples in training to enhance the learning ability. We tested PhaGCN on both simulated and real sequencing data. The results clearly show that our method competes favorably against available phage classification tools. \\
\textbf{Availability:} The source code of PhaGCN is available via: \href{https://github.com/KennthShang/PhaGCN}{https://github.com/KennthShang/PhaGCN}\\
\textbf{Supplementary information:} This paper was accepted by ISMB/ECCB 2021 and published on \textit{Bioinformatics} with DOI: 10.1093/bioinformatics/btab293
\end{abstract}

\section{Introduction}
\label{sec:intro}
Bacteriophages (or phages), which mainly infect bacteria, are among the most common and diverse biological entities in the biosphere \cite{mcgrath2007bacteriophage}. They regulate the actions of the ecosystem through killing, metabolic reprogramming, or gene transfer \cite{fernandez2018phage,hurwitz2016viral}. 
As a major agent of horizontal gene transfer between bacteria, phages can change the virulence of bacteria and indirectly cause human diseases. There are active studies that use phages for applications such as phage therapy \cite{loc2011pros}, disease diagnostics \cite{wang2004epitope,bazan2012phage}, and antimicrobial drug discovery \cite{liu2004antimicrobial}.  


Despite important functions of phages, our understanding of them is still very limited. Metagenomic sequencing, which allows us to obtain total genomic DNA directly from host-associated and environmental samples, has contributed significantly to new phage discovery \cite{bas2014phage,moon2018genomic,moon2020freshwater,moon2020viral}. In particular, metagenomic sequencing allows sequencing of uncultured dark matter of the microbial biosphere, which can contain a large amount of phages \cite{marine16}. The advancements of high-throughput sequencing, assembly, and contig scaffolding have led to phage-like contigs or genomes from different types of samples. According to the RefSeq database supported by the National Center for Biotechnology information (NCBI) , the number of identified phages changed from 1,468 in 2015 to 3,852 in 2020 in the RefSeq database, which is more than twice of increase. Despite the increase, known phages is just the tip of the iceberg of the virome on the planet \cite{santiago2019human}. How to automatically and accurately mine phages and assign their taxonomic groups from vast amount of sequencing data remains a challenging problem. 

There are two specific challenges for phage classification. First, the phages with known taxa are very limited. The International Committee on Taxonomy of Viruses (ICTV) is responsible for the official virus taxonomy and organizes viruses in order, family, subfamily, genera and species. Current ICTV classification procedures cannot catch up with new phage discovery. For example, one of the phage order named \textit{Caudovirales} has 3,691 reference genomes. However, there are more than 1,800 new \textit{Caudovirales} sequences found in 2020 that are unclassified into families. 
Limited labeled genomes pose challenges for both alignment-based and learning-based classification. Second, many phages in different taxa can share protein homologs, which adds ambiguity for alignment-based taxonomic classification. For example, more than 7,616 (\textasciitilde10\%) proteins in all annotated phage proteins are shared by phages in different families under \textit{Caudovirales}. In addition, more than 18,970 (\textasciitilde27\%) pairs of highly similar proteins  (E-value of BLASTP result < $10^{-50}$) are encoded by phages in different families. Therefore, using homology search alone can return ambiguous classification. 

In this work, we present a method that automates taxonomic classification for  contigs, which are the outputs of assembly. Although taxonomic classification can be conducted on both reads and contigs \cite{keegan2016mg}, recombination in viruses can make read-level taxonomic classification difficult. In addition, more distinctive features can be derived from contigs and thus can lead to improved classification accuracy.  Current metagenomic assembly tools, such as MEGAHIT \cite{li2015MEGAHIT}, have been extensively tested and can produce quality contigs from complex datasets. Thus, our tool accepts contigs as input. In order to address the aforementioned challenges, we developed a semi-supervised learning framework that incorporated the automatically learned features for each contig via a CNN, the protein sequence similarity, and the gene-sharing features between contigs/genomes. Both the unlabeled and labeled sequences were utilized for training in a graph convolutional neural network (GCN). We will demonstrate that the features from the unlabeled sequences (contigs) improve the learning ability and accuracy for phage classification. Below we summarize related work for phage classification.

\subsection{Related work}
\label{sec:relate}

Many attempts have been made for phage taxonomic classification. They can be roughly divided into two groups: alignment-based \cite{kristensen2013orthologous, aiewsakun2018evaluation, chibani2019classifying} and learning-based \cite{rohwer2002phage,jang2013phylogenomic,bolduc2017vcontact,jang2019taxonomic}. Alignment-based methods utilize either nucleotide-level or protein-level homology search between query contigs and reference genomes for assigning the taxon for the query. ClassiPhage \cite{chibani2019classifying} and Phage Orthologous Groups (POGs) \cite{kristensen2013orthologous} are two representative alignment-based phage classification tools. POGs extract taxon-specific marker genes and align query sequences against the marker genes using BLASTP. If there are statistical significant alignment for the contigs, the label of the best-aligned marker gene will be assigned to the contigs. ClassiPhage builds a profile Hidden Markov Model (pHMM) for each phage taxonomic group and apply HMM-based alignment for classification. 
There are two limitations with alignment-based method. First, as genes or proteins can be shared by different taxa, alignment-based method may lead to ambiguous label assignment or return a label with a higher rank using the lowest common ancestor (LCA) in the phylogenetic tree. Second, as phages are highly abundant and diverse, alignment-based methods are not able to assign taxa for new species that harbor novel proteins or lack quality alignments with the references.  For example, under the \textit{Caudovirales} order, 187,006 proteins are named as hypothetical proteins without known family labels. 13,382 proteins from phages released in 2020 do not have BLASTP results with the phages released before 2020. Thus, using only sequence similarity cannot provide ideal resolution. 


There are a number of learning-based tools for microbe classification such as the Naïve Bayes classifier \cite{wang2007naive} and CNN \cite{shang2020cheer}. They use either manually derived or automatically learned features to predict taxonomic labels for bacteria or RNA viruses.  The most relevant learning-based tool to phage classification is vConTACT 2.0 \cite{jang2019taxonomic}, which applies a graph clustering algorithm to assign labels for unknown contigs. 
In order to leverage gene organization conservation for phage classification, vConTACT utilizes a clustering algorithm to construct a gene-sharing network \cite{bolduc2017vcontact,jang2019taxonomic}. 
If the reference genomes and contigs are in the same cluster, the labels of the reference genomes will be assigned to those contigs. While these gene-sharing network methods present satisfactory performance on classification of complete genomes, the classification accuracy decreases as the length of the contigs becomes shorter. The decreased performance stems from the fact that short contigs do not contain many proteins and thus do not lead to valid edges in the gene-sharing network. As a result, the clustering algorithms fail to group contigs and reference genomes in the same cluster. Then, no labels will be assigned to these contigs. 


Given the enormous diversity of phages and the sheer amount of unlabeled phages, we formulate the phage classification problem as a semi-supervised learning problem. We choose the GCN as our learning model and combine the strength of both the alignment-based and the learning-based methods. First, we  utilize DIAMOND-derived sequence similarities between contigs and references  \cite{buchfink2015fast} to improve the edge construction process in the gene-sharing network. Second, to handle the situation that short contigs lack gene organization-related features, a CNN-based model is adopted to encode nucleotide information from the sequence. The GCN model allows us to utilize features from both labeled and unlabeled samples in training and thus lead to more accurate and sensitive phage classification. We compared our tool (named PhaGCN) with three state-of-the-art models specifically designed for phage classification: Phage Orthologous Groups (POG) \cite{kristensen2013orthologous}, vConTACT 2.0 \cite{jang2019taxonomic}, and ClassiPhage \cite{chibani2019classifying}. 
The experimental results demonstrated that PhaGCN outperforms other popular methods.


\section{Methods}

\begin{figure*}[h!]
    \centering
    \includegraphics[width=0.9\linewidth]{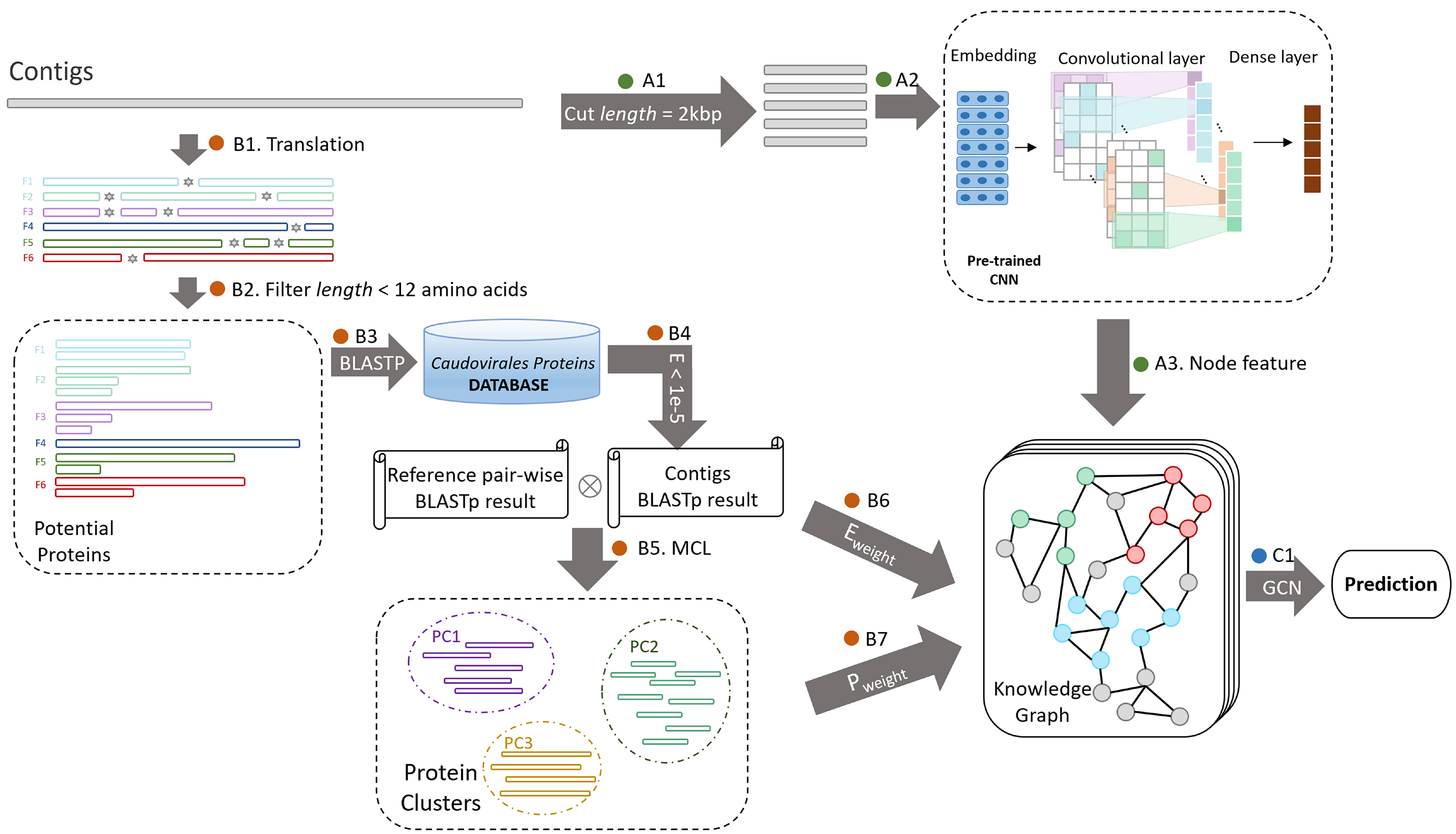}
    \caption{The pipeline of PhaGCN. A1: cut the contigs into 2kbp segments. A2: feature learning from the inputs using CNN. A3: construct nodes using encoded vectors. B1: contig translation using 6 reading frames. B2: filter short translations (12 amino acids). B3: align contigs against reference database using the DIAMOND BLASTP command. B4: choose the best translated frame for the BLASTP result. B5: use the BLASTP result to construct protein clusters. B6 and B7: define edges based on the sum of the $E_{weight}$ and $P_{weight}$. C1: construct the knowledge graph for GCN.}
    \label{fig:figure1}    
\end{figure*}

Semi-supervised learning is a machine learning approach that combines a small amount of labeled data with a large amount of unlabeled data during training. The main purpose of using the unlabeled data is to utilize their conjunction information with the labeled data to improve the classification accuracy. Because the number of reference (labeled) phage genomes is small and new (unlabeled) phage contigs are increasing quickly, we formulate the phage classification problem as a semi-supervised learning problem.

One of the semi-supervised learning approaches, named GCN, is based on deep learning. The basic idea of GCN is to apply a convolutional layer on a graph to utilize the features on non-Euclidean structure \cite{kipf2016semi}. The purpose of the graph convolutional layer is to automatically learn the topological features from the knowledge graph. Then, unlabeled samples/nodes can be represented as the weighted sum of their neighbor samples/nodes features. In biological data analyses, there exist many non-Euclidean structures such as protein topology graph on the supersecondary
structure, gene-sharing network, and diseases-gene relationship graph. GCN is expected to render high classification performance by employing the structural data. For example, GCN has shown promising results in finding the relationship between long non-coding RNAs and diseases \cite{zhao2020deeplgp, alam2020deep}. In phage classification, different phage genomes and contigs can share genes or proteins, which can be encoded in the graph of GCN. In addition, the nodes in GCN can embed automatically learned feature from nucleotide sequences. During the training, convolution is conducted for each node and its neighbors defined by the graph. The learned topological features will then be applied for classifying samples without labels.

The input to our GCN model is a knowledge graph. There are two key components in the knowledge graph: node encoding and edge construction. The node is a numerical vector learned from contigs using a CNN.
The edge encodes features from both the sequence similarity and the organization of genes.  Fig. \ref{fig:figure1} contains the major components for node and edge construction. To encode a sequence using a node, a pre-trained convolutional neural network (CNN) is adopted to capture features from the input DNA sequence (A1-A3). The CNN model is trained to convert proximate substrings into vectors of high similarity. The edge construction consists of several steps. We employ a greedy search algorithm to find the best BLASTP results (E-value less than 1e-5) between the 6-frame translations of the contigs and the database (B1-B4). Then the Markov clustering algorithm (MCL) is applied to generate protein clusters from the BLASTP result (B5) \cite{jang2019taxonomic}. Based on the results of BLASTP (sequence similarity) and MCL (shared proteins), we define the edges between sequences (contigs and reference genomes) using two metrics: $P_{weight}$ and $E_{weight}$ (B6-B7). By combining the node’s features and edges (C1), we construct the knowledge graph  and feed it to the GCN to classify new phage contigs.

\subsection{Using CNN to encode input sequences in the knowledge graph}

CNN can automatically learn motif-related features for sequence classification \cite{alipanahi2015predicting, seo2018deepfam}. Although CNN can be directly applied to phage classification, our experiments will show that using CNN alone cannot render the best classification performance. Thus, we only train the CNN for encoding input contigs. 

\begin{figure*}[h!]
    \centering
    \includegraphics[width=0.8\linewidth]{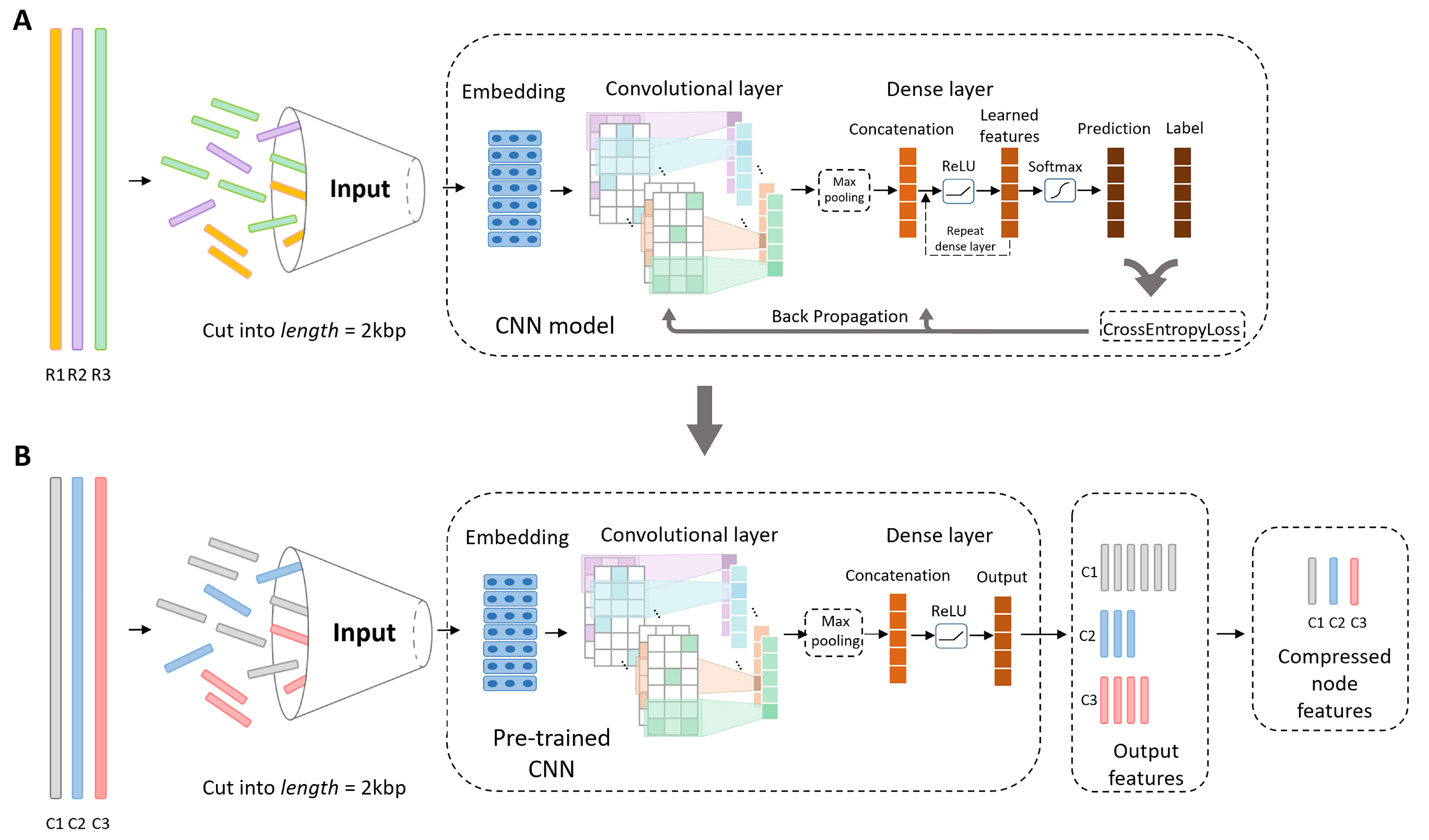}
    \caption{The structures of the CNN model in PhaGCN for training (A) and encoding (B). R1,R2,R3 are the reference genomes used for training. C1,C2,C3 are the contigs that need to be encoded. In train mode, sequences will be fed to CNN to update parameters during back propagation. In encoding mode, The pre-trained CNN will be used to encode the sequences into numerical vectors. Then these vectors will be adopted as node features in the knowledge graph.}
    \label{fig:figure2}    
\end{figure*}

As shown in Fig. \ref{fig:figure2}, there are two slightly different network structures in the CNN for ``train mode'' and ``encoding mode'', respectively. In the train mode (Fig. \ref{fig:figure2}(A)), we use the reference database to train the CNN model. In the encoding mode (Fig. \ref{fig:figure2}(B)), the output of the first dense layer in the pre-trained CNN will be used to encode sequences into numerical vectors.

\noindent \textbf{Train mode:} Because the CNN model can only handle fixed length input, all the inputs will be cut into 2kbp segments with user-specified stride value (default 50). The segment has the same label as the underlying genome according to the ICTV taxonomic classification.

The CNN model contains three different parts: embedding layer, convolutional layer, and dense layer. The embedding layer is used to convert the DNA sequence into numerical inputs for convolution. There are two major methods for the embedding layer: one-hot embedding and skip-gram embedding \cite{mikolov2013distributed}. As shown in our previous work of using CNN for classifying RNA viruses \cite{shang2020cheer}, the skip-gram based embedding can improve CNN's learning ability. Thus, in this work, we implemented a skip-gram embedding layer that can map proximate k-mers into highly similar vectors. We trained the embedding layer using k-mers and their neighboring (proximate) k-mers so that the embedding layer can learn their adjacent relationship. Specifically, in order to train the embedding layer, we use a 3-mer at position $i$ as input and 3-mers located at $i+j$ as output, where $-m \leq j \leq m$. m is the hyperparameter that can be specified for the skip-gram model. We employ 100 hidden units in the embedding layer to encode the 3-mers and the output of the embedded vector has 100 dimensions. Thus, each 2kbp segment is converted into embedded matrix $M \in \mathbb{R}^{2,000 \times 100}$.

\begin{equation}\label{cnn1}
Z_{i}(M, w_{conv})= ReLU( {\textstyle \sum_{j=1}^{n_{conv}}}w _{conv}^{j}*M[i:i+d_{1}-1][1:d_{2}]+b)
\end{equation}

\begin{equation}\label{cnn2}
H^{(0)} = Maxpool(Z(M, w_{conv}))
\end{equation}

\begin{equation}\label{cnn3}
H^{(l+1)} = ReLU(H^{(l)},w^{(l)})
\end{equation}

\begin{equation}\label{cnn4}
output\ of\ train\ mode = SoftMax(H^{(2)},w^{(2)})
\end{equation}

Then, the embedded matrix $M$ will be fed into the convolutional layer. Eq. \ref{cnn1} is the convolution function. b is the bias term; $d_{1}$ and $d_{2}$ are the filter sizes. Since the embedded vector has 100 dimensions, $d_{2} = 100$. $M[i:i+d_{1}-1][1:d_{2}]$ defines a 2D window size of $d_{1}\times d_{2}$ of the embedded matrix $M$. ReLU is the activation function. The convolutional filters $w_{conv}$ contain $n_{conv}$ 2D matrices and $w_{conv}^{j}$ is the $j$-th filter.  We applied filters repeatedly to each possible window of the input embedded matrix to produce a feature map. Then the dense layer is applied to compress the features captured by the convolutional layer as shown in Eq. \ref{cnn2} and Eq. \ref{cnn3}. First, max-pooling (Eq. \ref{cnn2}) is applied to the feature map to capture the most useful information from the convolutional layer. Second, we use two dense layers with ReLU activation function (Eq. \ref{cnn3}) to learn and compress the feature map. $H^{(l)}$ is the feature map in hidden layer $l$ and $w^{(l)}$ is the weight parameters in the $l$-th hidden layer. Since we only has two dense layers, $l \in \{0, 1\}$. Finally, the SoftMax function (Eq. \ref{cnn4}) is adopted to generate the prediction. As shown in Fig. 2, in the train mode, we employ CrossEntropyLoss to calculate the error between prediction and real label and backpropagate the loss to update the parameters in the model. The detailed parameters are listed in our Github repository.

\noindent \textbf{Encoding mode:}
After training the CNN model, we utilize the pre-trained parameters to convert contigs into numerical vectors. The main difference in the encoding mode is that we only use the output of the first dense layer as the learned feature rather than using the SoftMax function for prediction.
Eq. \ref{cnn5} shows the equation to convert $x$ (an input 2kbp segment) into the output of the first dense layer. If a contig is cut into multiple segments of length 2kbp, we will conduct vector addition for all the segments' outputs and divide it by the number of segments. Thus, contigs of different lengths are always converted into vectors of the same size (determined by the units of the dense layer, default 512). 


\begin{equation}
\label{cnn5}
    Out(x) = ReLU (pool(Z(M, w_{conv}), w^{(0)})
\end{equation}

\subsection{Construction of the edge in GCN}
The edges connect nodes that are likely in the same taxonomic group. We define the edge by incorporating both the number of shared protein clusters and also the average protein similarity between two sequences. Intuitively, if two sequences share a large number of common protein clusters with high similarity, they tend to belong to the same taxa. In order to quantify the significance of two sequences sharing $c$ common proteins, we first define protein clusters. A pair of proteins from two sequences is called a shared protein if they are in the same protein cluster.  

\subsubsection{Construction of the protein cluster}
\label{sec:dataset}
We follow the idea in \cite{bolduc2017vcontact, jang2019taxonomic} to construct protein clusters. We start by extracting proteins from all sequences. For the genomes in the reference database, proteins are downloaded from NCBI RefSeq. For the input contigs, DNA sequences are translated into amino acid sequences using 6 reading frames. 
We employ DIAMOND to conduct all-against-all pairwise alignment between contigs’ 6-frame translations and proteins encoded by the genomes. If there are multiple alignments for different reading frames of a contig, only the best frame is kept. Then we create a weighted graph where the node is a protein sequence in the contig or genome and the edge represents an alignment with E-value less than a threshold. The edge weight is the E-value. Then protein clusters are subsequently identified using the Markov clustering algorithm (MCL). Finally, clusters that contain at least two proteins will be kept. 

\subsubsection{Definition of the edges}
\label{d_edges}
The edge is defined by computing two metrics: $P_{weight}$ and $E_{weight}$. 
$P_{weight}$ is adopted to calculate the expected number of sequences sharing at least an observed number of common proteins (i.e. $c$ proteins). 
Following vConTACT \cite{bolduc2017vcontact}, by assuming that each of the $n$ protein clusters has the same chance to be chosen, we compute the probability that any two sequences containing $a$ and $b$ protein clusters share at least $c$ clusters in Eq. \ref{edge1}.  
Eq. \ref{edge2} then computes the expected number of sequence pairs with at least $c$ common proteins out of $\binom{N}{2}$ sequence pairs, where $N$ is the number of sequences (contigs and reference genomes). With increase of $c$, $P$ in Eq. \ref{edge1} becomes small enough to return a positive $P_{weight}$. 
\begin{equation}
    \label{edge1}
    P(y \ge c) =  {\textstyle \sum_{i=c}^{min(a,b)}} \frac{\binom{a}{i}\binom{n-a}{b-i}}{\binom{n}{b}}
\end{equation}

\begin{equation}
\label{edge2}
    P_{weight} = -log(P(y \ge c) \times \binom{N}{2})
\end{equation}


While $P_{weight}$ is used to evaluate whether two sequences share a significant number of common proteins, $E_{weight}$ is adopted to calculate the sequence similarity using alignments' E-values.
For two sequences $A$ and $B$ with $N_{c}$ shared proteins, we first define $S(A,B)$ in Eq. \ref{edge3}, which is the arithmetic mean of the E-values of $N_{c}$ alignments. $S(A,B)$ has a small value only when all the shared proteins have significant E-values, which helps reduce false edge construction for short contigs.
For each genome $A$, $S(A,A')$ is ranked for all A's adjacent nodes $A'$. Users can decide how many edges to keep by specifying a threshold. By default, we only keep the top 5 edges for each genome $A$. For all the kept edges, $E_{weight}=S(A,B)$.

\begin{equation}
\label{edge3}
    S(A,B) = \left\{\begin{matrix}
  0 ,& if\ no\ alignment\ result\\
   {-log(\frac{\textstyle \sum_{i=0}^{N_{c}}e_{value}(i)}{N_{c}}}) ,& otherwise
\end{matrix}\right.
\end{equation}

\noindent \textbf{Edges in the knowledge graph:}
The final edge in the knowledge graph is defined based on the sum of  $P_{weight}$ and $E_{weight}$.  An edge is defined when the sum is above a threshold $\tau$, which is 1 by default (Eq. \ref{edge5}). It connects two sequences with enough common proteins of high similarity. Usually, as long contigs share more proteins with the reference genome database, $P_{weight}$ tends to big enough for creating edges between the long contigs and the knowledge graph. However, short contigs have fewer shared proteins and thus we use $E_{weight}$ to examine whether the shared proteins have significant similarities with references for creating an edge. If a contig has no edge connecting to the knowledge graph, PhaGCN will not output a prediction. Only contigs in the knowledge graph will be fed to the GCN for training and prediction.

\begin{equation}
\label{edge5}
   Edge = \left\{\begin{matrix}
  1,& if\ P_{weight} + E_{weight} > \tau \\
  0,& otherwise
\end{matrix}\right.
\end{equation}

\subsection{The GCN model}
After constructing the knowledge graph, we train a GCN to assign labels for all unlabeled contigs. 

\begin{equation}
    \label{gcn1}
    H^{(l+1)} = ReLU (\tilde{D}^{-\frac{1}{2}} \tilde{G} \tilde{D}^{\frac{1}{2}}H^{(l)}W^{(l)})
\end{equation}
\begin{equation}
    \label{gcn2}
    Out = SoftMax(H^{(2)}W^{(2)})
\end{equation}

\begin{figure}[h!]
    \centering
    \includegraphics[width=0.8\linewidth]{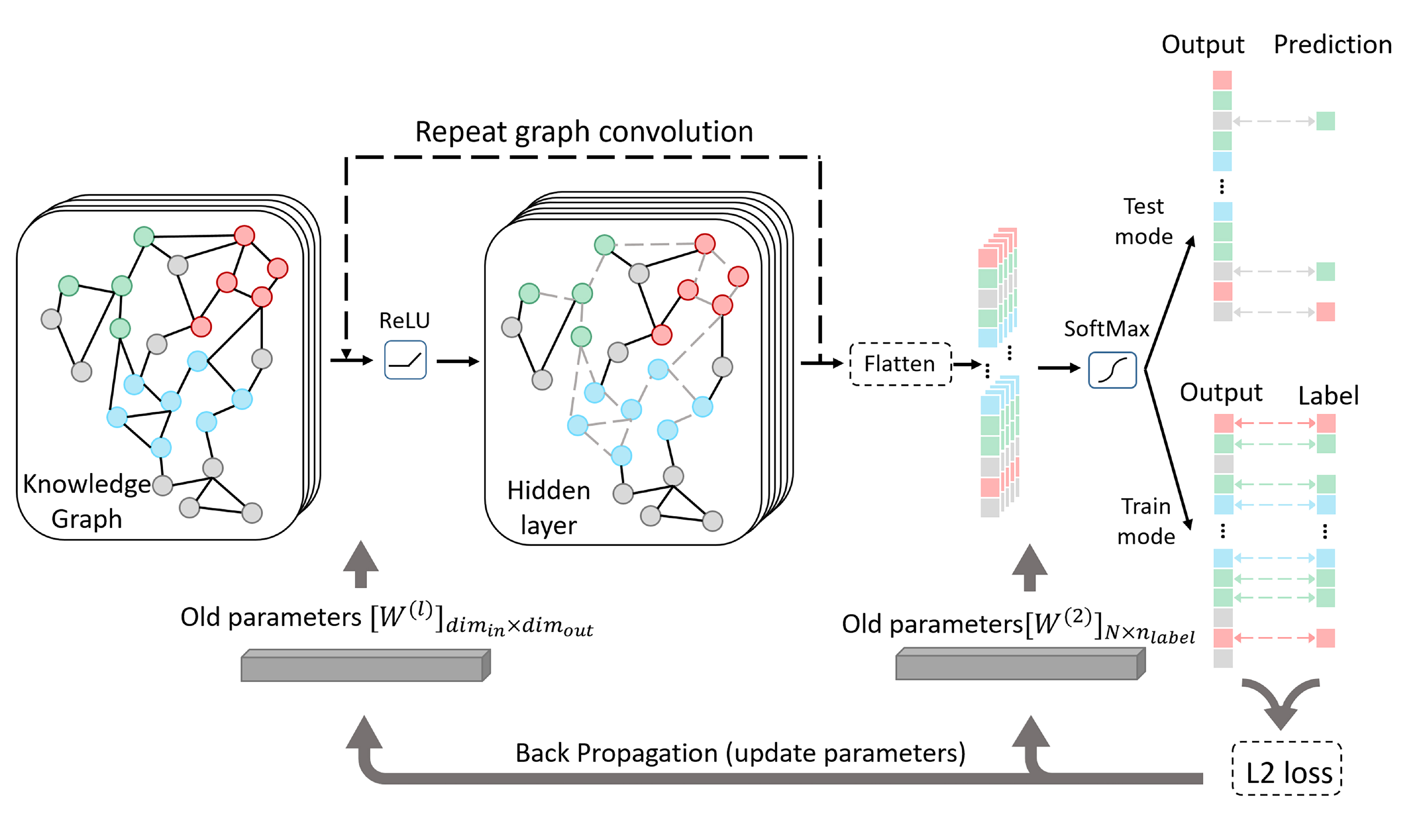}
    \caption{The structure of the GCN model in PhaGCN. Both feature from labeled (green, red, and blue color) and unlabled samples (grey color) will be used in training and prediction. In train mode, only labeled samples will be utilize to calculate the loss and update the parameters. In test mode, the SoftMax function will be applied to generate prediction for unlabeled samples.}
    \label{fig:figure3}  
\end{figure}

The basic concept of graph convolutional layer is shown in Eq. \ref{gcn1}. Suppose we have N sequences (nodes) in the knowledge graph. $G$ is the $\mathbb{R}^{N \times N}$ adjacency matrix of the knowledge graph and $I_{N}$ is an $\mathbb{R}^{N \times N}$ identity matrix.  $\tilde{G}$ is calculated with $\tilde{G} = G + I_{N}$. $\tilde{D}$ is the $\mathbb{R}^{N \times N}$ diagonal matrix calculated with $D_{ij} = {\textstyle \sum_{j}} \tilde{G}_{ij}$. $H^{(l)}$ is the feature map in the $l$-th hidden layer; $H^{(0)}$ is the node feature matrix; and $W^{(l)}$ is a matrix of weight parameters. After the graph convolutional layer, we apply a dense layer and use the SoftMax function to calculate the output matrix $Out \in \mathbb{R}^{N \times n_{label}}$ (Eq. \ref{gcn2}). Because we have two graph convolutional layers and one dense layer in our model, $l \in \{0, 1\}$. The output dimension $n_{label}$ is decided by the number of classes in the database. As shown in Fig. \ref{fig:figure3},  only the labeled samples will be used to calculate the loss in the training process. We adopt L2 loss to calculate the error between prediction and the labeled samples and back propagate the loss to update the weight parameters. After training the GCN model, we freeze the parameters and use the SoftMax value of the unlabeled samples to assign their labels in the test mode. The detailed parameters are listed in our Github repository.

\section{Result}
\label{sec:exp}
\subsection{Data and performance metrics}
We demonstrate the performance of PhaGCN on classifying contigs in families under \textit{Caudovirales}, which is an order containing the majority of known phages from RefSeq (95.8\% of total phage reference genomes). We downloaded the \textit{Caudovirales} reference genomes from the NCBI RefSeq database. As shown in Table \ref{tab:dataset}, there are 3,639 genomes from 8 different families. As the lower ranks contain few genomes in each group, we focus on family-level classification in the experiment. 

\begin{table}[h!]
\centering
\begin{tabular}{ll}
\hline
Name              & Number of genomes \\ \hline
\textit{Ackermannviridae}  & 63                \\
\textit{Autographiviridae} & 378               \\
\textit{Demerecviridae}    & 87                \\
\textit{Drexlerviridae}    & 112               \\
\textit{Herelleviridae}    & 136               \\
\textit{Myoviridae}        & 775               \\
\textit{Podoviridae}       & 337               \\
\textit{Siphoviridae}      & 1805              \\ \hline
\end{tabular}
\caption{\centering{8 families under \textit{Caudovirales}.}}
\label{tab:dataset}
\end{table}

\subsubsection{Data and experiment design}
PhaGCN was tested on both simulated and real sequencing data. For the simulated data, we applied two different methods to generate contigs with known labels: 1. randomly sample contigs from the reference genome; 2. simulate reads with  ART-Illumina \cite{huang2012art} and run MEGAHIT \cite{li2015MEGAHIT} to assemble contigs. After validating PhaGCN on simulated data with known ground truth, we downloaded two real sequencing datasets from NCBI SRA and evaluated PhaGCN on assembled contigs. As phages are highly abundant in marine environment samples \cite{marine16}, we tested PhaGCN on virus-like contigs from 71 metagenomic data sets that are sequenced from oyster. We recorded macro-accuracy, macro-recall, and macro-precision for each experiment (Eq.\ref{m1}, Eq. \ref{m2}, and Eq. \ref{m3}) when the ground truth can be derived. 
$N_{class}$ is the total number of classes. $TP$ is the True positive, $TN$ is the True negative, and $FN$ is the false negative. $Acc_{i}$ is the accuarcy of class $i$. Except for CNN, each tool can output either a family label or no label at all (no prediction). For each class, if its positive samples have no predictions, they are counted as $FN$. If its negative samples have no predictions, they are counted as $TN$. As macro-average will compute each metric independently for each class and then take the average, these metrics treat all classes equally. 

\begin{equation}
    \label{m1}
    Acc_{macro} = \frac{ {\textstyle \sum_{i=0}^{N_{class}}} Acc_{i} }{N_{class}}  
\end{equation}
\begin{equation}
    \label{m2}
    Precision_{macro} = \frac{ {\textstyle \sum_{i=0}^{N_{class}}}Precision_{i} }{N_{class}} = \frac{ {\textstyle \sum_{i=0}^{N_{class}}}\frac{TP_{i}}{TP_{i}+FP_{i}}}{N_{class}}
\end{equation}
\begin{equation}
    \label{m3}
    Recall_{macro} = \frac{ {\textstyle \sum_{i=0}^{N_{class}}}Recall_{i} }{N_{class}} =\frac{ {\textstyle \sum_{i=0}^{N_{class}}} \frac{TP_{i}}{TP_{i}+FN_{i}}}{N_{class}}
\end{equation}

The main purpose of PhaGCN is to classify new phages that do not have reference genomes in the training data. When the training and testing data share common genomes, high accuracy may be attributed to memorization rather than learning. 
Thus, in all the experiments conducted using PhaGCN, we use genome-masking, meaning that the genomes in the testing data will be removed from the training data so that they do not share any genomes. The test contigs thus represent novel phages. 

We compare our results with three representative and widely used pipelines: vConTACT 2.0, Phage Orthologous Groups (POGs), and ClassiPhage. In addition, as CNN itself can conduct classification, we also compared with the CNN model we trained for PhaGCN.


\subsection{Experiments using simulated contigs}
\label{sec:exp1}
In this experiment, we randomly chose 20 genomes from each family as the testing species. The remaining genomes were used as the training set. Thus the testing data contain 160 genomes from all 8 families under \textit{Caudovirales}. Then, we sampled contigs from each test genome by generating random starting positions. 
To estimate the impact of contig length on PhaGCN, we generated contigs in three length ranges: [4kbp, 8kbp], [8kbp, 12kbp], and [12kbp, 16kbp]. Each contig is generated using a random start position and a random length within each range. For each of the length range, we generated 10 contigs. Thus we have 1,600 contigs for testing. Finally, we repeated this experiment for three times and recorded the average performance of the three experiments. In total, for each length range, 4,800 contigs from 480 genomes were tested. We also recorded the results of using complete genomes as the test data. To have a fair comparison, we applied the same method to construct the training set (or reference database) and the test set for vConTACT 2.0, POGs, and ClassiPhage. 


\begin{figure}[h!]
    \centering
    \includegraphics[width=0.5\linewidth]{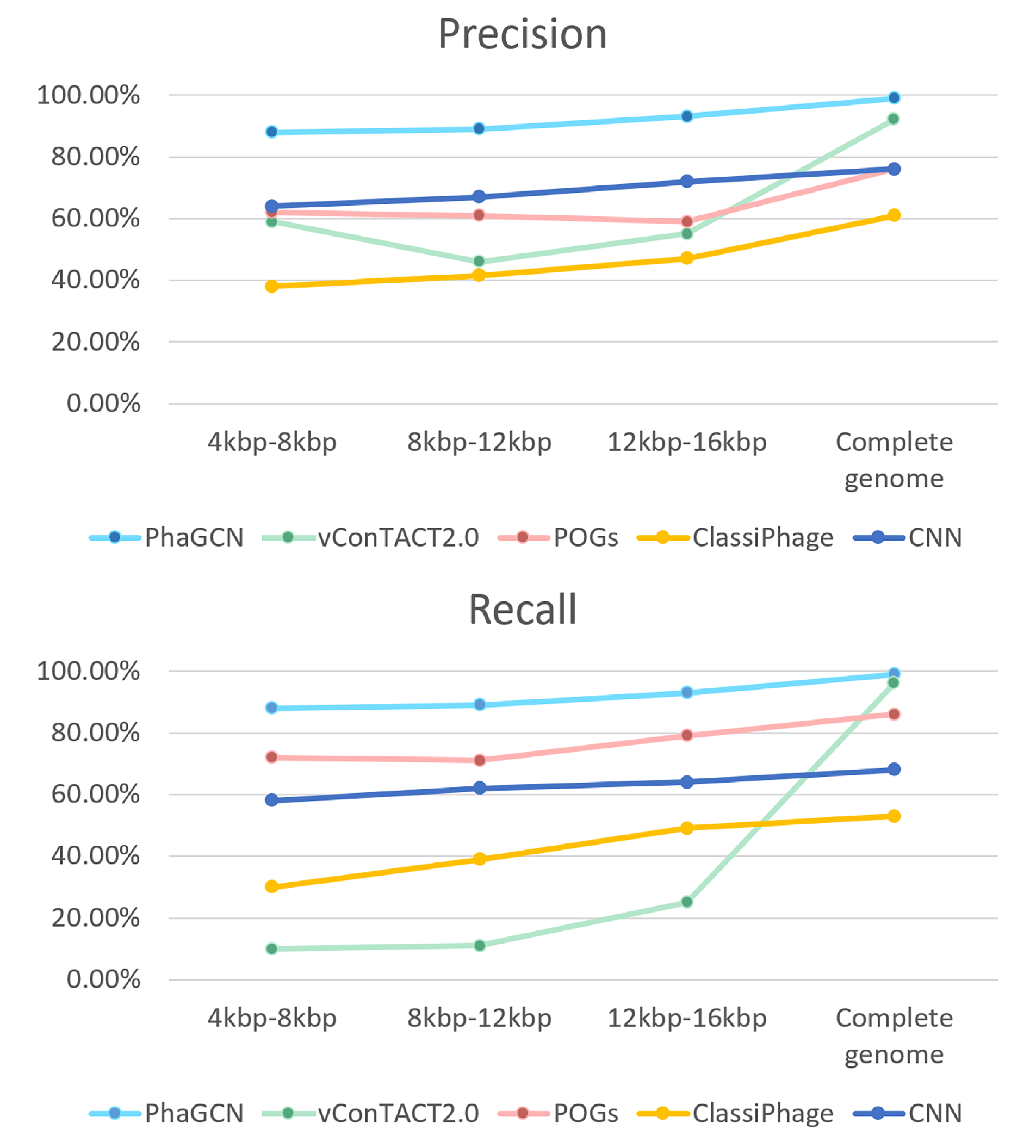}
    \caption{The precision and recall (Eq. \ref{m2} and Eq. \ref{m3}) of PhaGCN, vConTACT2.0, ClassiPhage, POGs, and the CNN model in PhaGCN on simulated contigs, which are randomly sampled from phage genomes. X-axis: the length range of contigs. For each length range, there are 1,600 randomly sampled contigs  from 20 genomes of 8 families. The reported performance is averaged on three such sets of contigs for each length range. }
    \label{fig:figure4}    
\end{figure}

Fig. \ref{fig:figure4} shows that PhaGCN outperforms other state-of-the-art tools across different length range. With the increase of contig length, the performance of all pipelines increases. This is expected because longer contigs contain more proteins, which can lead to better classification performance. We also evaluated the classification performance of only using the CNN model in PhaGCN. Both the recall and precision of using  GCN is better than using only CNN, showing that the knowledge graph enhances the learning ability of the model. In addition, the classification performance of PhaGCN is stable with the change of the contig length, making it useful for classifying short contigs.

\begin{figure}[h!]
    \centering
    \includegraphics[width=0.5\linewidth]{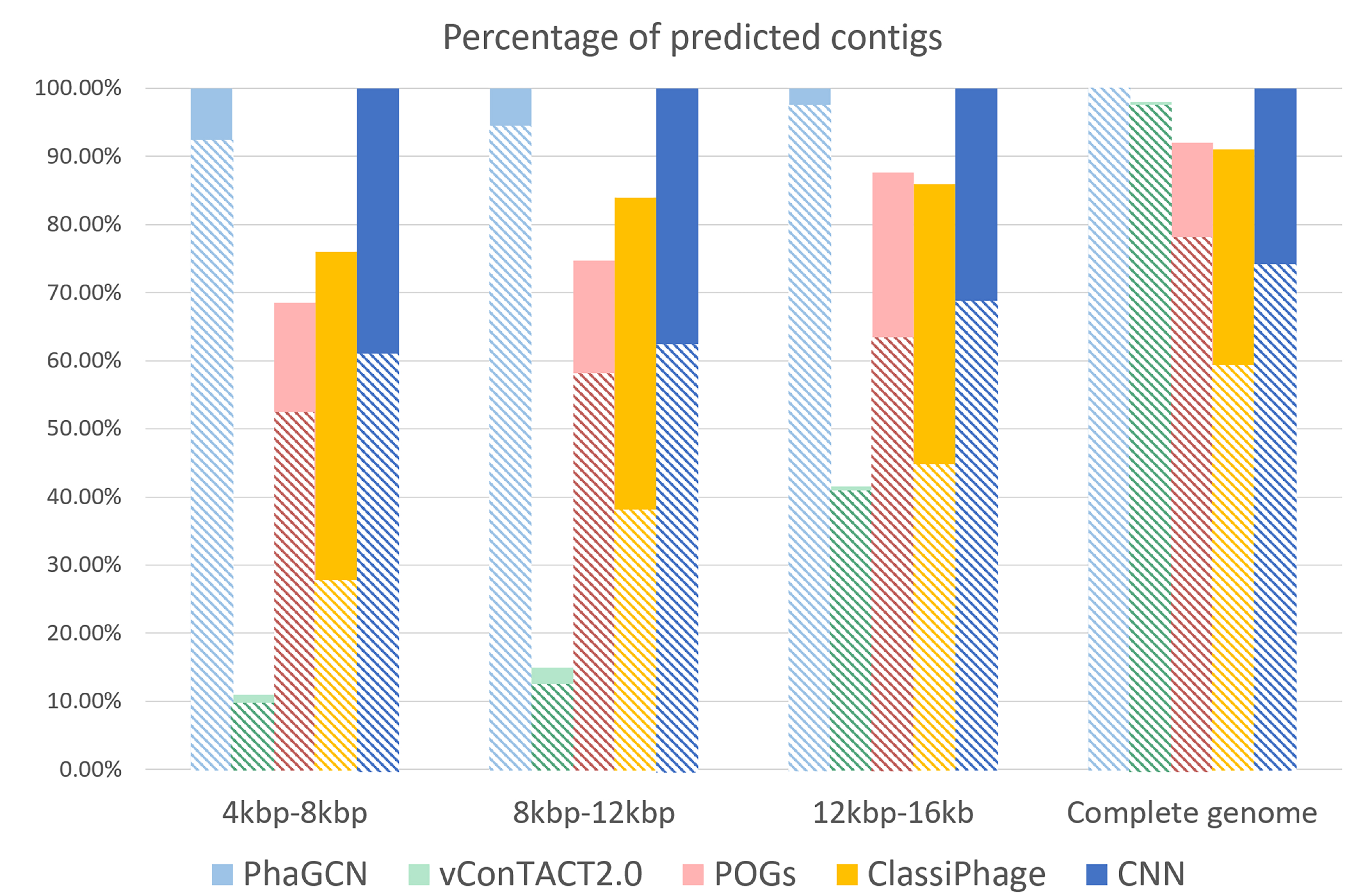}
    \caption{The percentage of classified contigs and classification accuracy (Eq.\ref{m1}) of each model on simulated contigs, which are randomly sampled from phage genomes. Each bar shows the percentage of classified contigs. The solid part shows the misclassification rate and the bottom part with patterns represents the macro-accuracy.}
    \label{fig:figure5}   
\end{figure}

Fig. \ref{fig:figure5} shows the classification accuracy of all tools on randomly sampled contigs. The bar height shows the percentage of predicted contigs. In addition, each bar is divided into two parts. The top part (solid) is the misclassification rate while the bottom part with patterns corresponds to the macro-accuracy for the classified contigs. The result shows that PhaGCN always has the largest number of predicted contigs with the highest classification accuracy. 

Although vConTACT 2.0 can assign the correct labels to most of the predicted contigs as shown in Fig. \ref{fig:figure5}, it only generated predictions for a small number of the contigs in two families (\textit{Podoviridae} and \textit{Siphoviridae}). For families without any prediction, the precision is 0. Thus, the macro-precision of vConTACT is small. 


\subsection{Experiments using simulated reads}
\label{sec:exp2}

In this experiment, we downloaded all newly released \textit{Escherichia coli} phages under \textit{Caudovirales} in 2020 from NCBI RefSeq and used them as the testing species (a total of 99 species are downloaded). And we downloaded all \textit{Caudovirales} phages released in 2019 from NCBI RefSeq and used them as the training set. Consequently, these newly released \textit{Escherichia coli} phages can be treated as unknown phages for our model. Then we applied ART-Illumina to simulate reads from the testing sequences. The parameters used for generating reads are -p, -l 150, -ss HS25, -f 20, -m 200 and -s 10. The output contains 150bp paired-end reads simulated under HiSeq 2500. We mixed all the simulated reads in one dataset and run MEGAHIT to assemble them. In order to quantify the performance of different tools, we determine the correct label of contigs by aligning them against the test genomes using BLAST in glocal mode. Only contigs of length above 2kbp with taxon-specific alignment results and query coverage $>$ 85\% will be kept. As a result, a total of 301 contigs were used as input for comparison. 

\begin{figure}[h!]
    \centering
    \includegraphics[width=0.5\linewidth]{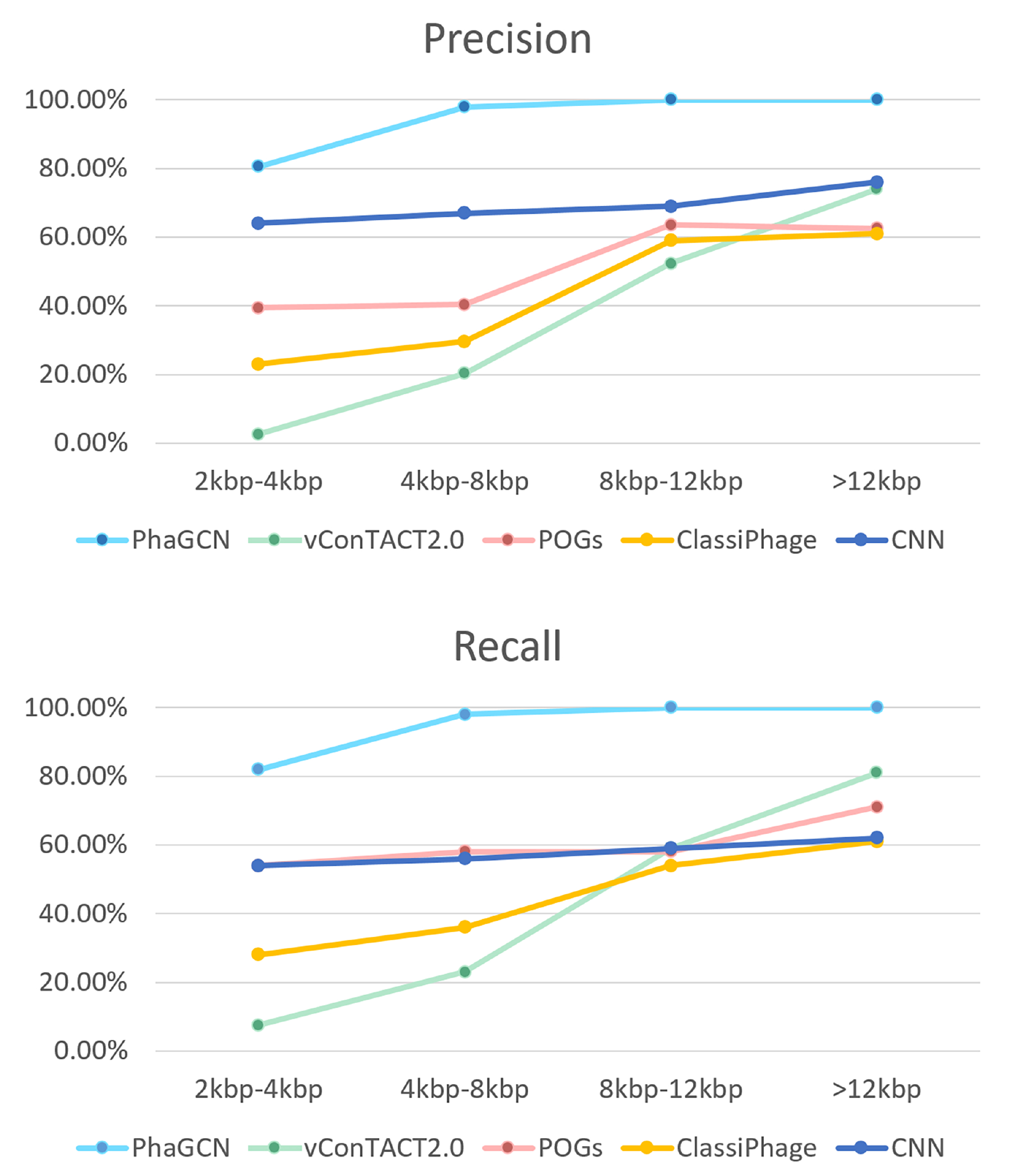}
    \caption{The precision and recall (Eq. \ref{m2} and Eq. \ref{m3}) of PhaGCN, vConTACT2.0, ClassiPhage, POGs and the CNN model in PhaGCN on contigs that are assembled using MEGAHIT from simulated reads. 301 contigs assembled from simulated reads of 99 species are used as inputs. The numbers of contigs for each length range are 101, 107, 51, and 42, respectively.} 
    \label{fig:figure6}    
\end{figure}

\begin{figure}[h!]
    \centering
    \includegraphics[width=0.5\linewidth]{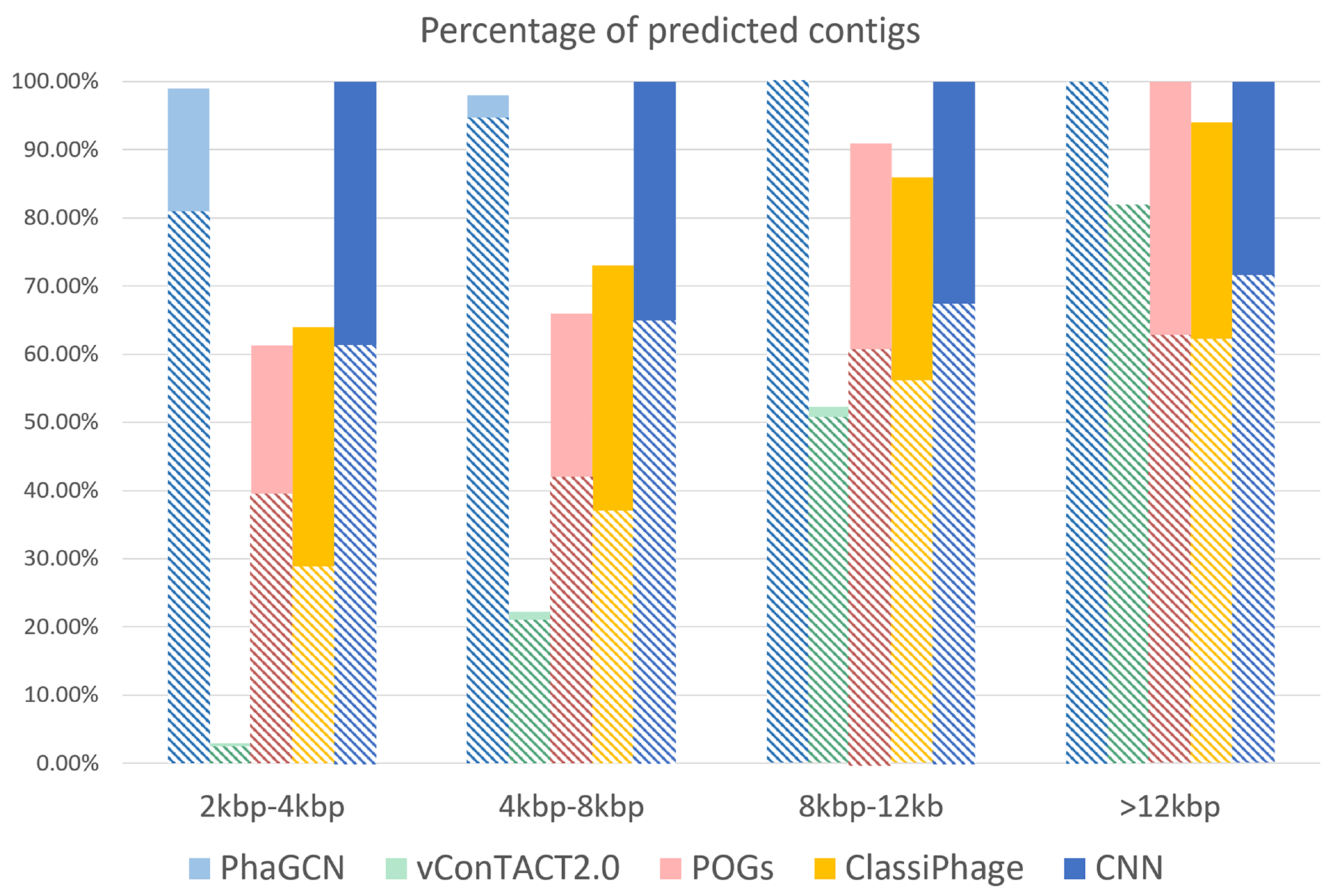}
    \caption{The percentage of classified contigs and classification accuracy (Eq.\ref{m1}) of each model on contigs that are assembled using MEGAHIT from simulated reads. Each bar shows the percentage of classified contigs. The solid part shows the misclassification rate and the bottom part with patterns represents the macro-accuracy. } 
    \label{fig:figure7}    
\end{figure}

As shown in Fig. \ref{fig:figure6} and Fig. \ref{fig:figure7}, PhaGCN outperforms other tools across different length, which is consistent with the conclusion in Section \ref{sec:exp1}. We also find that when the length of the contigs becomes shorter ([2kbp, 4kbp]), PhaGCN can still achieve over 80\% accuracy. 
Although the reads simulated by ART-Illumina contain sequencing error, the performance of PhaGCN still achieves high accuracy (100\% for contigs over 8kbp). Because we only tested newly released \textit{Escherichia coli} phages in 2020, the classification performance of all methods were slightly better than the results in Section \ref{sec:exp1}.

\subsection{Running time comparison}
The most resource demanding component in PhaGCN is the sequence alignment. We used it to produce the protein clusters by conducting pairwise alignments between contigs and reference sequences. PhaGCN produces protein clusters for each set of input contigs by assuming that they may contain novel proteins. In addition, alignment is also conducted for defining the edge in the knowledge graph. Table \ref{tab:time} shows the average elapsed time of classifying 100 contigs for each tool.  PhaGCN is not the fastest program. Optimization can be applied to reduce the number of pairwise alignments. For example, we can produce a database of protein clusters and reduce the number of pairwise alignments.

\begin{table}[h!]
\centering
\begin{tabular}{lllll} \hline
Program                       & PhaGCN & vConTACT 2.0 & ClassiPhage & POGs \\ \hline
Elapsed time(min/100 contigs) & 12     & 32           & 4           & 7 \\ \hline
\end{tabular}
\caption{The average elapsed time to predict labels of 100 contigs for each method. All the methods are run on Intel\textsuperscript{\textregistered} Xeon\textsuperscript{\textregistered} Gold 6258R CPU with 8 cores.}
\label{tab:time}
\end{table}

\subsection{Experiments on real sequencing data}
\label{sec:exp3}

\begin{table}[h!]
\centering
\begin{tabular}{ll}
\hline
SRR12949983               & SRR13132427                      \\ \hline
\textit{Escherichia phage C5}      & \textit{Escherichia phage V18}            \\
\textit{Salmonella phage C2}       & \textit{Escherichia virus FV3}            \\
\textit{Serratia phage Pila}       & \textit{Escherichia virus JES2013}        \\
\textit{Escherichia virus E112}    & \textit{Escherichia phage CEC\_Kaz\_2018} \\
\textit{Escherichia virus ECML134} & \textit{Escherichia phage SECphi18}       \\
\textit{Escherichia virus T4}      & \textit{Escherichia phage vB\_EcoS\_PNS1}\\
\hline
\end{tabular}
\caption{Reference species in SRR12949983 and SRR13132427. These species are shown in the associated taxonomic analysis at NCBI SRA. We use them as the ground truth to assign labels to the assembled contigs.}
\label{tab:table2}

\end{table}

\begin{table}[h!]
\centering
\begin{tabular}{lllllll}
\hline
\multicolumn{3}{c}{SRR12949983}                                                  & \multicolumn{4}{c}{SRR13132427}                                                         \\ \hline
\multicolumn{7}{c}{\textbf{No. of contigs assembled by MEGAHIT (\textgreater{}2kbp)}}                                                                                    \\ \hline
\multicolumn{3}{l}{24 (including 1 bacterial and 1 unknown)}                     & \multicolumn{4}{l}{20 (including 1 bacterial)}                                          \\ \hline
\textbf{Phage family}              & \multicolumn{2}{l}{\textbf{No. of contigs}} & \multicolumn{2}{l}{\textbf{Phage family}} & \multicolumn{2}{l}{\textbf{No. of contigs}} \\ \hline
\textit{Autographiviridae}                  & \multicolumn{2}{l}{5}                       & \multicolumn{2}{l}{\textit{Myoviridae}}            & \multicolumn{2}{l}{14}                      \\
\textit{Myoviridae}                         & \multicolumn{2}{l}{17}                      & \multicolumn{2}{l}{\textit{Siphoviridae}}          & \multicolumn{2}{l}{5}    \\  \hline                
\end{tabular}
\caption{Contigs assembled by MEGAHIT  and their family labels derived from their alignments against the species in Table \ref{tab:table2}. }
\label{tab:table3}
\end{table}

In this experiment, we searched for real sequencing data that contain \textit{Caudovirales} at NCBI SRA and downloaded two datasets, SRR12949983 and SRR13132427. Then we used MEGAHIT to assemble reads into contigs on these two datasets separately. To quantify the performance of phage classification on these two datasets, we used the provided read-level taxonomic analysis by NCBI SRA as the ground truth. The phages in the two datasets provided by NCBI SRA are listed in Table \ref{tab:table2}. We used the same method introduced in Section \ref{sec:exp2} to label the contigs and removed all these genomes in Table \ref{tab:table2} from the reference database before training PhaGCN. The contigs in the test data are listed in Table \ref{tab:table3}.

\begin{figure}[h!]
    \centering
    \includegraphics[width=0.6\linewidth]{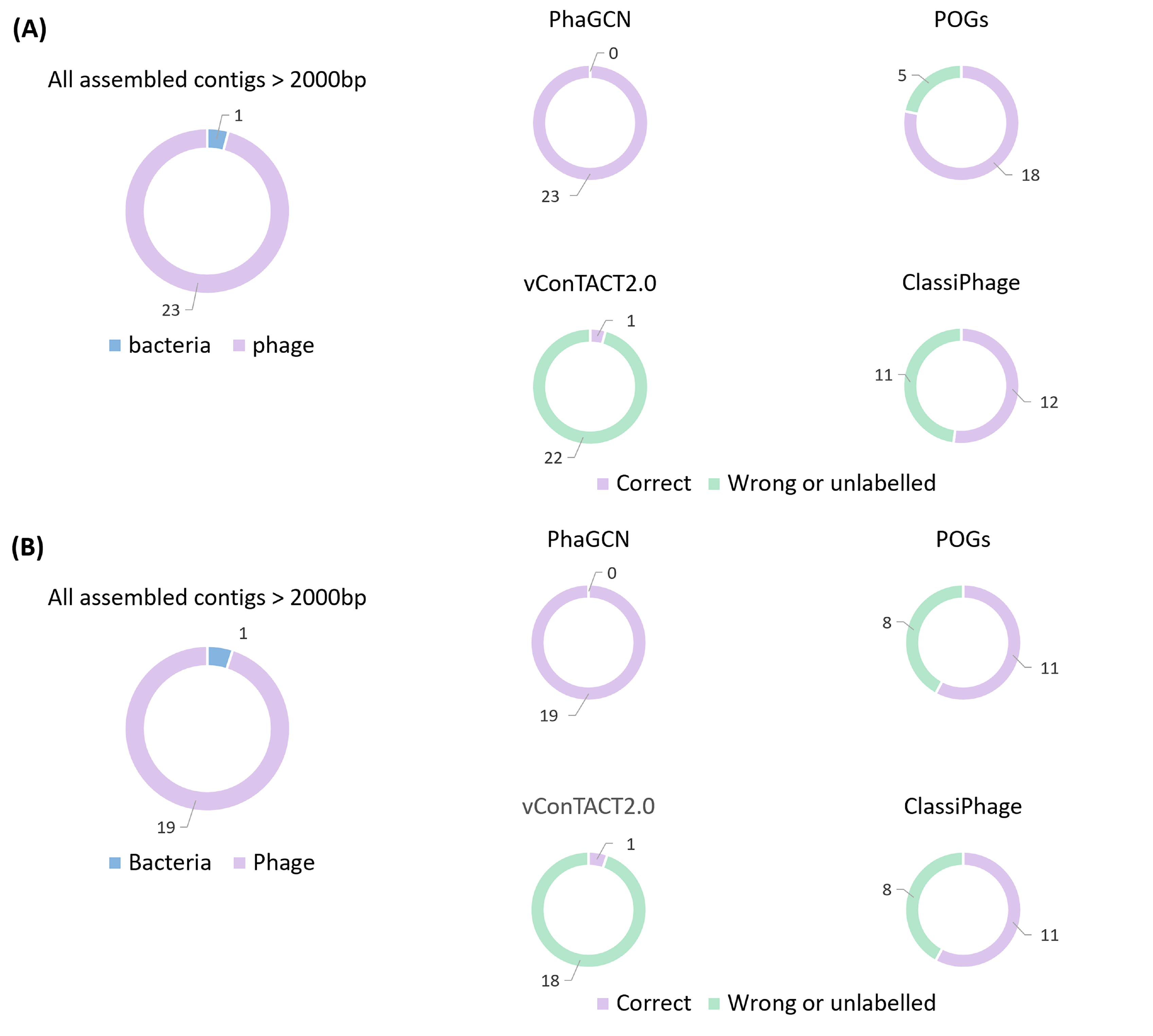}
    \caption{The classification result of PhaGCN, vConTACT 2.0, POGs, and ClassiPhage on SRR12949983(A) and SRR13132427(B).}
    \label{fig:figure8}    
\end{figure}

We compared the phage labels assigned by PhaGCN, vConTACT, POGs, and ClassiPhage on the assembled contigs. Because the real sequencing data might contain bacteria sequences, we run DeepVirFinder \cite{ren2020identifying} to reject the contigs that belong to bacteria. As shown in Fig. \ref{fig:figure8}, there is only one bacterial contig in each dataset. All other contigs (43 contigs) were fed into the four tools for classification.

Fig. \ref{fig:figure8} reveals that PhaGCN achieves better classification performance than vConTACT, POGs, and ClassiPhage. PhaGCN has 100\% accuracy in both datasets. Many contigs could not be classified by vConTACT 2.0 (unlabeled in Fig. \ref{fig:figure8}). POGs and ClassiPhage are able to assign labels for all contigs but a number of them have wrong family labels.

In the dataset SRR121949983, there is one contig lacking ground truth because the contig cannot be aligned to any reference genome in Table \ref{tab:table2}. By extending the reference database to the Nucleotide Collection (nr/nt) database, BLAST shows that this contig belongs to a phage in \textit{Siphoviridae}. We input this contig to the four tools and only PhaGCN assigned the correct label. It is worth noting that the reference genome is not in the training data of PhaGCN. The performance of alignment-based approaches heavily relies on the reference database while PhaGCN can learn the features for classifying new contigs.

\begin{figure}[h!]
    \centering
    \includegraphics[width=0.5\linewidth]{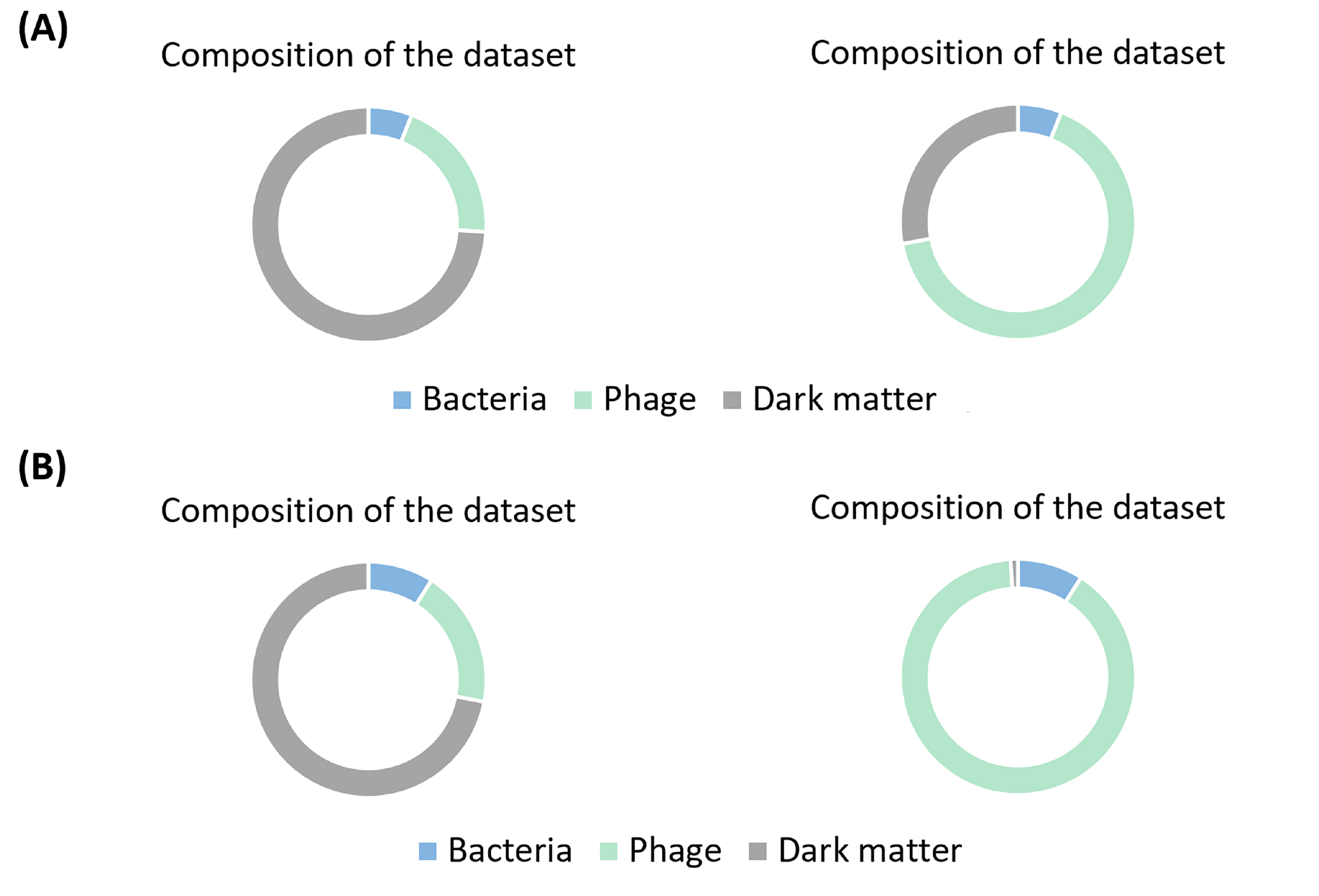}
    \caption{The composition analysis of SRR12949983(A) and SRR13132427(B). Left: composition analysis published at NCBI SRA. Right: composition analysis presented by PhaGCN.}
    \label{fig:figure9}    
\end{figure}

Furthermore, we showed the composition of these two datasets before and after using PhaGCN in Fig. \ref{fig:figure9}. The results revealed that our model can greatly improve the composition analysis for the dark matter. Thus, PhaGCN can benefit metagenomic analysis.

\subsection{Phage classification in contigs produced from oyster metagenomic data}
After validating PhaGCN on simulated datasets and two real sequencing datasets, we applied PhaGCN to contigs assembled from  metagenomic data of oyster samples. The samples were collected by the co-author Dr. Jiang between April, 2016 to July, 2019 from various sites along the coast of South China Sea. Metagenomic sequencing was conducted from samples in the gill, visceral mass, and mantle tissues of oyster using viral-like particle enrichment and protocols in \cite{wei2018detection, jing2018isolation}. There are about 2.5 billions of raw reads from 71 libraries.

\begin{table*}[h!]
\resizebox{\textwidth}{6mm}{
\begin{tabular}{cccccccccc} \hline
            & \textit{Ackermannviridae} & \textit{Autographiviridae} & \textit{Demerecviridae} & \textit{Drexlerviridae} & \textit{Herelleviridae} & \textit{Myoviridae} & \textit{Podoviridae} & \textit{Siphoviridae} & unclassified \\ \hline
vConTACT    & 0                         & 0                          & 0                       & 0                       & 0                       & 50                  & 116                  & 102                   & 22698        \\
PhaGCN & 150                       & 1727                       & 301                     & 74                      & 32                      & 5880                & 2682                 & 6173                  & 5767        \\ \hline
\end{tabular}}
\caption{Prediction results of PhaGCN and vConTACT for contigs produced from the oyster metagenomic data.}
\label{tab:table4}
\end{table*}

After applying standard quality control and MEGAHIT with the default setting, there are about 3,375,091 contigs of length above 500bp. After removing contigs that can be aligned to bacteria, archaea, eukaryota, we kept 22,966 contigs with length above 4,000bp as input to PhaGCN. Of them, 17,199 contigs can be assigned to \textit{Caudovirales} by PhaGCN.  

When users apply composition analysis for their samples, precision is important for generating valid hypothesis. Because vConTACT 2.0 has higher accuracy for what they can predict than ClassiPhage and POGs based on our experiments, we compared PhaGCN with vConTACT 2.0 in this experiment and summarized the results in Table \ref{tab:table4}. 
Although we don't have the ground truth for this large-scale metagenomic sequencing data, the numbers of predicted contigs are consistent with the results of experiments on simulated and real sequencing datasets. The contigs vConTACT 2.0 can classify are significantly less than PhaGCN. 
\textasciitilde 74.8\% contigs are predicted by PhaGCN while \textasciitilde 1.1\% contigs are predicted by vConTACT 2.0. The contigs predicted by vConTACT is a subset of PhaGCN. We found that the clustering algorithm in vConTACT failed to group contigs and reference genomes in the same cluster. There exist many clusters containing only unlabeled contigs and thus, no labels will be assigned to these contigs.
Also, PhaGCN can identify more families from the dataset. Because the classification accuracy of PhaGCN is more than 92\% when the length of contigs is over 4kbp, the result can provide useful family-level composition analysis for the oyster metagenomic data.  

\subsection{Extension of PhaGCN}
\begin{figure}[h!]
    \centering
    \includegraphics[width=0.6\linewidth]{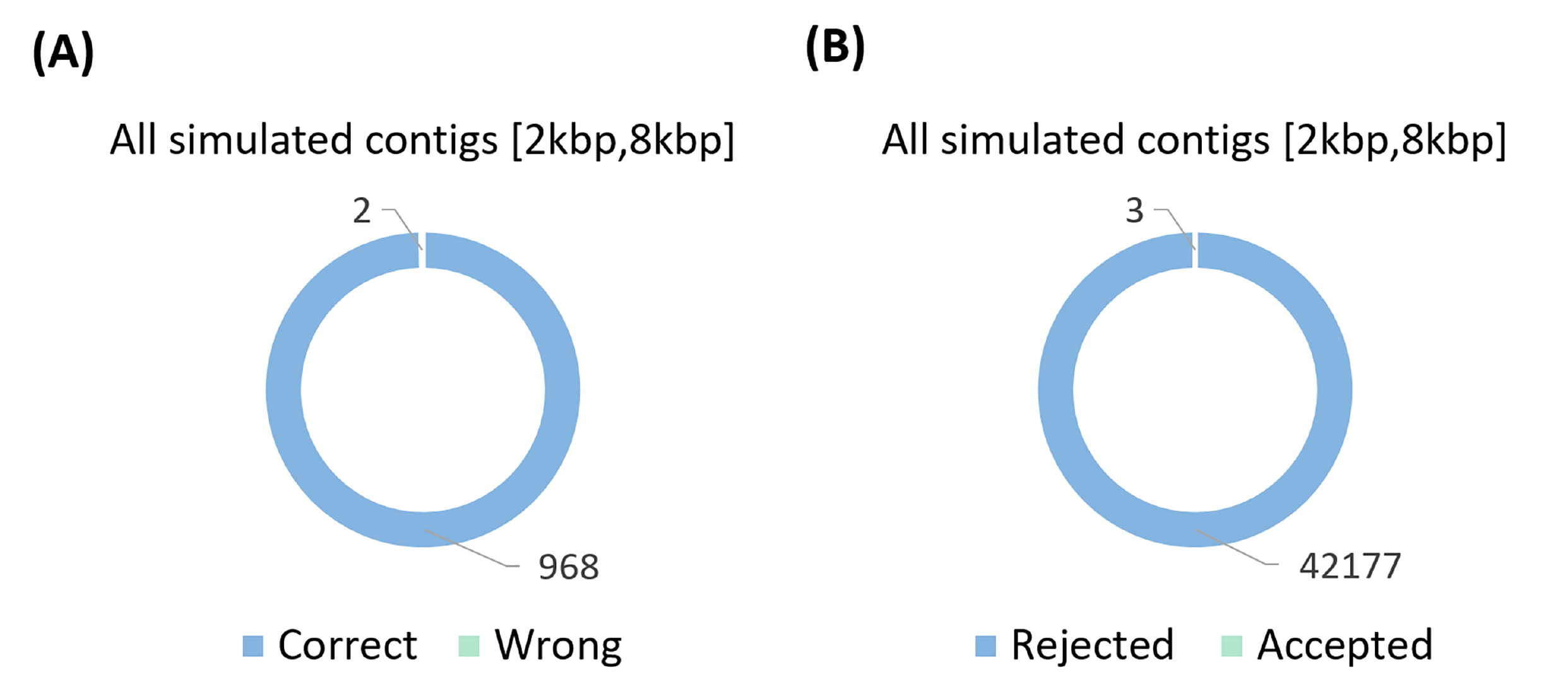}
    \caption{(A): the classification result of three added families: \textit{Rudiviridae}, \textit{Microviridae}, and \textit{Inoviridae}. (B): PhaGCN can reject non-\textit{Caudovirales} phages.}
    \label{fig:figure10}   
\end{figure}

As \textit{Caudovirales} is the order with the most number of sequenced phages from RefSeq, we validated PhaGCN on classifying families in this order. But PhaGCN can be conveniently extended to other taxa. We extended PhaGCN by adding families that contain at least 10 genomes. Using this criterion, three families (\textit{Rudiviridae}, \textit{Microviridae}, and \textit{Inoviridae}) were added.  We used the leave-one-genome-out method to choose testing genomes from these three families and use the same method introduced in Section \ref{sec:exp1} to generate contigs. The classification results in Fig. \ref{fig:figure10}(A) show that PhaGCN can correctly classify almost all of the contigs from the extended families. 

Using knowledge graph enables PhaGCN to detect targeted phage families, which is useful for applications where only some phages are of interest.  Specifically, phages that are not in the training families usually won't form edges with the nodes in the graph and thus will not be mis-classified into \textit{Caudovirales}. We validated the detection ability of PhaGCN by testing whether PhaGCN can reject contigs that do not belong to \textit{Caudovirales}. We downloaded 4,218 phage genomes that do not belong to \textit{Caudovirales} from RefSeq according to the ICTV taxonomic affiliation information. For each genome, we apply the same method introduced in Section \ref{sec:exp1} to generate 10 contigs for each of them. Thus, a total of 42,180 contigs were tested. As shown in Fig. \ref{fig:figure10}(B), only 3 of them are accepted (predicted) by PhaGCN. This experiment demonstrates that PhaGCN can be applied for targeted phage detection.

\section{Discussion}
As shown in the experiments, the performance of alignment-based approaches, such as POGs and BLAST, heavily relies on the reference database. The ambiguous hits or lack of reference genomes for highly divergent or novel phages can decrease the classification accuracy. Existing learning-based tools like vConTACT 2.0 cannot achieve good performance on short contigs. In this work, we demonstrate that PhaGCN can render better performance for novel phage classification. The major improvement of our method stems from combined strength of the reference-based model and the learning-based model using the knowledge graph: the nodes contain automatically learned features from nucleotide sequences and the edges are created by protein-based alignment. Then the semi-supervised GCN is applied on the knowledge graph to utilize both labeled and unlabeled data for training. 

Although PhaGCN has greatly improved phage contig classification, we have several goals to optimize or extend PhaGCN in our future work. First, we simplified the edge weight computation by assuming that all protein clusters can be chosen with the same probability. We will investigate whether incorporating protein cluster size and the cluster's entropy in label distribution can render more accurate edges. Second, PhaGCN can reject non-relevant phages with high accuracy. Thus we will extend it to phage detection and compare it with DeepVirFinder. Third, although we have demonstrated that PhaGCN can be easily extended to more families, it is still hard to predict classes with only a few training samples (less than 10). We will incorporate relevant learning methods to improve the classification accuracy for small families. Finally, we will explore whether we can incorporate bacteria in our knowledge network for phage host detection. This can be used to further validate the classification results on the oyster metagenomic data.

\section*{Funding}
This work was supported by Hong Kong RGC GRF 9042828 and HKIDS (9360163) and NSF of China (31972847).

\bibliographystyle{unsrt}  
\bibliography{references}  

\begin{thebibliography}{10}

\bibitem{mcgrath2007bacteriophage}
Stephen McGrath, D~van Sinderen, et~al.
\newblock {\em {Bacteriophage: genetics and molecular biology.}}
\newblock Caister Academic Press, 2007.

\bibitem{fernandez2018phage}
Luc{\'\i}a Fern{\'a}ndez, Ana Rodr{\'\i}guez, and Pilar Garc{\'\i}a.
\newblock {Phage or foe: an insight into the impact of viral predation on
  microbial communities}.
\newblock {\em The ISME journal}, 12(5):1171--1179, 2018.

\bibitem{hurwitz2016viral}
Bonnie~L Hurwitz and Jana~M U’Ren.
\newblock {Viral metabolic reprogramming in marine ecosystems}.
\newblock {\em Current opinion in microbiology}, 31:161--168, 2016.

\bibitem{loc2011pros}
Catherine Loc-Carrillo and Stephen~T Abedon.
\newblock {Pros and cons of phage therapy}.
\newblock {\em Bacteriophage}, 1(2):111--114, 2011.

\bibitem{wang2004epitope}
Lin-Fa Wang and Meng Yu.
\newblock {Epitope identification and discovery using phage display libraries:
  applications in vaccine development and diagnostics}.
\newblock {\em Current drug targets}, 5(1):1--15, 2004.

\bibitem{bazan2012phage}
Justyna Bazan, Ireneusz Ca{\l}kosi{\'n}ski, and Andrzej Gamian.
\newblock {Phage display—A powerful technique for immunotherapy: 1.
  Introduction and potential of therapeutic applications}.
\newblock {\em Human vaccines \& immunotherapeutics}, 8(12):1817--1828, 2012.

\bibitem{liu2004antimicrobial}
Jing Liu, Mohammed Dehbi, Greg Moeck, Francis Arhin, Pascale Bauda, Dominique
  Bergeron, Mario Callejo, Vincent Ferretti, Nhuan Ha, Tony Kwan, et~al.
\newblock {Antimicrobial drug discovery through bacteriophage genomics}.
\newblock {\em Nature biotechnology}, 22(2):185--191, 2004.

\bibitem{bas2014phage}
Bas~E Dutilh, Noriko Cassman, Katelyn McNair, Savannah~E Sanchez, Genivaldo G~Z
  Silva, Lance Boling, Jeremy~J Barr, Daan~R Speth, Victor Seguritan, Ramy~K
  Aziz, Ben Felts, Elizabeth~A Dinsdale, John~L Mokili, and Robert~A Edwards.
\newblock {A highly abundant bacteriophage discovered in the unknown sequences
  of human faecal metagenomes}.
\newblock {\em Nature Communications}, 5:4498, 2014.

\bibitem{moon2018genomic}
Kira Moon, Ilnam Kang, Suhyun Kim, Sang-Jong Kim, and Jang-Cheon Cho.
\newblock {Genomic and ecological study of two distinctive freshwater
  bacteriophages infecting a Comamonadaceae bacterium}.
\newblock {\em Scientific reports}, 8(1):1--9, 2018.

\bibitem{moon2020freshwater}
Kira Moon, Jeong~Ho Jeon, Ilnam Kang, Kwang~Seung Park, Kihyun Lee, Chang-Jun
  Cha, Sang~Hee Lee, and Jang-Cheon Cho.
\newblock {Freshwater viral metagenome reveals novel and functional phage-borne
  antibiotic resistance genes}.
\newblock {\em Microbiome}, 8:1--15, 2020.

\bibitem{moon2020viral}
Kira Moon, Suhyun Kim, Ilnam Kang, and Jang-Cheon Cho.
\newblock {Viral metagenomes of {Lake Soyang}, the largest freshwater lake in
  South Korea}.
\newblock {\em Scientific Data}, 7(1):1--6, 2020.

\bibitem{marine16}
Blanca Perez~Sepulveda, Tamsin Redgwell, Branko Rihtman, Frances Pitt, David~J.
  Scanlan, and Andrew Millard.
\newblock {Marine phage genomics: the tip of the iceberg}.
\newblock {\em FEMS Microbiology Letters}, 363(15), 06 2016.
\newblock fnw158.

\bibitem{santiago2019human}
Tasha~M Santiago-Rodriguez and Emily~B Hollister.
\newblock {Human virome and disease: high-throughput sequencing for virus
  discovery, identification of phage-bacteria dysbiosis and development of
  therapeutic approaches with emphasis on the human gut}.
\newblock {\em Viruses}, 11(7):656, 2019.

\bibitem{keegan2016mg}
Kevin~P Keegan, Elizabeth~M Glass, and Folker Meyer.
\newblock {MG-RAST, a metagenomics service for analysis of microbial community
  structure and function}.
\newblock In {\em Microbial environmental genomics (MEG)}, pages 207--233.
  Springer, 2016.

\bibitem{li2015MEGAHIT}
Dinghua Li, Chi-Man Liu, Ruibang Luo, Kunihiko Sadakane, and Tak-Wah Lam.
\newblock {MEGAHIT: an ultra-fast single-node solution for large and complex
  metagenomics assembly via succinct de Bruijn graph}.
\newblock {\em Bioinformatics}, 31(10):1674--1676, 2015.

\bibitem{kristensen2013orthologous}
David~M Kristensen, Alison~S Waller, Takuji Yamada, Peer Bork, Arcady~R
  Mushegian, and Eugene~V Koonin.
\newblock {Orthologous gene clusters and taxon signature genes for viruses of
  prokaryotes}.
\newblock {\em Journal of bacteriology}, 195(5):941--950, 2013.

\bibitem{aiewsakun2018evaluation}
Pakorn Aiewsakun, Evelien~M Adriaenssens, Rob Lavigne, Andrew~M Kropinski, and
  Peter Simmonds.
\newblock {Evaluation of the genomic diversity of viruses infecting bacteria,
  archaea and eukaryotes using a common bioinformatic platform: steps towards a
  unified taxonomy}.
\newblock {\em The Journal of general virology}, 99(9):1331, 2018.

\bibitem{chibani2019classifying}
Cynthia~Maria Chibani, Anton Farr, Sandra Klama, Sascha Dietrich, and Heiko
  Liesegang.
\newblock {Classifying the unclassified: a phage classification method}.
\newblock {\em Viruses}, 11(2):195, 2019.

\bibitem{rohwer2002phage}
Forest Rohwer and Rob Edwards.
\newblock The {Phage Proteomic Tree}: a genome-based taxonomy for phage.
\newblock {\em Journal of bacteriology}, 184(16):4529--4535, 2002.

\bibitem{jang2013phylogenomic}
Ho~Bin Jang, Fernand~F Fagutao, Seong~Won Nho, Seong~Bin Park, In~Seok Cha,
  Jong~Earn Yu, Jung~Seok Lee, Se~Pyeong Im, Takashi Aoki, and Tae~Sung Jung.
\newblock {Phylogenomic network and comparative genomics reveal a diverged
  member of the $\phi$kz-related group, marine Vibrio phage $\phi$JM-2012}.
\newblock {\em Journal of virology}, 87(23):12866--12878, 2013.

\bibitem{bolduc2017vcontact}
Benjamin Bolduc, Ho~Bin Jang, Guilhem Doulcier, Zhi-Qiang You, Simon Roux, and
  Matthew~B Sullivan.
\newblock {vConTACT: an iVirus tool to classify double-stranded DNA viruses
  that infect Archaea and Bacteria}.
\newblock {\em PeerJ}, 5:e3243, 2017.

\bibitem{jang2019taxonomic}
Ho~Bin Jang, Benjamin Bolduc, Olivier Zablocki, Jens~H Kuhn, Simon Roux,
  Evelien~M Adriaenssens, J~Rodney Brister, Andrew~M Kropinski, Mart Krupovic,
  Rob Lavigne, et~al.
\newblock {Taxonomic assignment of uncultivated prokaryotic virus genomes is
  enabled by gene-sharing networks}.
\newblock {\em Nature biotechnology}, 37(6):632--639, 2019.

\bibitem{wang2007naive}
Qiong Wang, George~M Garrity, James~M Tiedje, and James~R Cole.
\newblock {Naive Bayesian classifier for rapid assignment of rRNA sequences
  into the new bacterial taxonomy}.
\newblock {\em Applied and environmental microbiology}, 73(16):5261--5267,
  2007.

\bibitem{shang2020cheer}
Jiayu Shang and Yanni Sun.
\newblock {CHEER: hierarCHical taxonomic classification for viral mEtagEnomic
  data via deep leaRning}.
\newblock {\em Methods}, 2020.

\bibitem{buchfink2015fast}
Benjamin Buchfink, Chao Xie, and Daniel~H Huson.
\newblock Fast and sensitive protein alignment using diamond.
\newblock {\em Nature methods}, 12(1):59--60, 2015.

\bibitem{kipf2016semi}
Thomas~N Kipf and Max Welling.
\newblock {Semi-supervised classification with graph convolutional networks}.
\newblock {\em arXiv preprint arXiv:1609.02907}, 2016.

\bibitem{zhao2020deeplgp}
Tianyi Zhao, Yang Hu, Jiajie Peng, and Liang Cheng.
\newblock {DeepLGP: a novel deep learning method for prioritizing lncRNA target
  genes}.
\newblock {\em Bioinformatics}, 36(16):4466--4472, 2020.

\bibitem{alam2020deep}
Tanvir Alam, Hamada~RH Al-Absi, and Sebastian Schmeier.
\newblock {Deep Learning in LncRNAome: Contribution, Challenges, and
  Perspectives}.
\newblock {\em Non-coding RNA}, 6(4):47, 2020.

\bibitem{alipanahi2015predicting}
Babak Alipanahi, Andrew Delong, Matthew~T Weirauch, and Brendan~J Frey.
\newblock {Predicting the sequence specificities of DNA-and RNA-binding
  proteins by deep learning}.
\newblock {\em Nature biotechnology}, 33(8):831--838, 2015.

\bibitem{seo2018deepfam}
Seokjun Seo, Minsik Oh, Youngjune Park, and Sun Kim.
\newblock {DeepFam: deep learning based alignment-free method for protein
  family modeling and prediction}.
\newblock {\em Bioinformatics}, 34(13):i254--i262, 2018.

\bibitem{mikolov2013distributed}
Tomas Mikolov, Ilya Sutskever, Kai Chen, Greg Corrado, and Jeffrey Dean.
\newblock {Distributed representations of words and phrases and their
  compositionality}.
\newblock {\em arXiv preprint arXiv:1310.4546}, 2013.

\bibitem{huang2012art}
Weichun Huang, Leping Li, Jason~R Myers, and Gabor~T Marth.
\newblock {ART: a next-generation sequencing read simulator}.
\newblock {\em Bioinformatics}, 28(4):593--594, 2012.

\bibitem{ren2020identifying}
Jie Ren, Kai Song, Chao Deng, Nathan~A Ahlgren, Jed~A Fuhrman, Yi~Li, Xiaohui
  Xie, Ryan Poplin, and Fengzhu Sun.
\newblock {Identifying viruses from metagenomic data using deep learning}.
\newblock {\em Quantitative Biology}, pages 1--14, 2020.

\bibitem{wei2018detection}
Hong-Ying Wei, Sheng Huang, Tuo Yao, Fang Gao, Jing-Zhe Jiang, and Jiang-Yong
  Wang.
\newblock {Detection of viruses in abalone tissue using metagenomics
  technology}.
\newblock {\em Aquaculture Research}, 49(8):2704--2713, 2018.

\bibitem{jing2018isolation}
Jiang Jingzhe and Wei Hongying.
\newblock {Isolation of Viral Like Particles (VLP) from Tissues of Molluscs}.
\newblock {\em protocols.io}, 2018.

\end{thebibliography}

\end{document}